\documentclass[11pt,twoside]{article}
\usepackage[top=1in, bottom=1in, left=1.25in, right=1.25in]{geometry}
\usepackage{latexsym}
\usepackage{amsmath}
\usepackage{amssymb}
\usepackage{txfonts}
\usepackage{graphicx}
\usepackage{epsfig}
\allowdisplaybreaks

\def\ra{\rangle}
\def\la{\langle}
\def\be{\begin{equation}}
\def\ee{\end{equation}}
\def\ba{\begin{array}}
\def\ea{\end{array}}

\begin{document}
\title
 {\Large \bf Bell inequality for multipartite qubit quantum system and the maximal violation}
\author
{ Ming Li$^1$ and Shao-Ming Fei$^{2,3}$}
\date{}
\maketitle
\begin{center}
\begin{minipage}{140mm}

\small $~^{1}$ {\small Department of Mathematics, School of Science,
China University of Petroleum, 266555 Qingdao }

\small $~^{2}$ {\small School of Mathematical Sciences, Capital
Normal University, 100048 Beijing}

{\small $~^{3}$ Max-Planck-Institute for Mathematics in the
Sciences, 04103 Leipzig}

\vspace{2ex}

 {\bf\small Abstract~} {\small We present a
set of Bell inequalities for multi-qubit quantum systems.  These Bell inequalities are shown to be
able to detect multi-qubit entanglement better than previous Bell inequalities such as WWZB
ones. Computable formulas are presented for calculating the maximal violations of
these Bell inequalities for any multi-qubit states.

{\bf Keywords~} Bell inequality; Entanglement; Separability.

\vfill }
\end{minipage}
\end{center}

\bigskip
Based on the Einstein, Podolsky and Rosen (EPR) gedanken experiment,
Bell presented an inequality obeyed by any local hidden variable
theory \cite{bell}. It turns out that this inequality and its
generalized forms are satisfied by all separable quantum states,
but may be violated by pure entangled states and some mixed quantum
states \cite{vbell1,vbell2,vbell3,chenjingling,liprl,yu}. Thus Bell
inequalities are of great importance both in understanding the
conceptual foundations of quantum theory and in investigating
quantum entanglement. Bell inequalities are also closely related to
certain tasks in quantum information processing, such as building
quantum protocols to decrease communication complexity \cite{dcc}
and providing secure quantum communication \cite{scc1}. Due to their
significance, Bell inequalities have been generalized from two
qubit case, such as the Clauser-Horne-Shimony-Holt (CHSH)
inequality \cite{chsh} to the N-qubit case, such as the
Mermin-Ardehali-Belinskii-Klyshko (MABK) inequality
\cite{mabk,chenka}, and to arbitrary d-dimensional (qudit)
systems such as the Collins-Gisin-Linden-Masser-Popescu (CGLMP)
inequality \cite{cglmp}. However, except for some special cases such
as bipartite pure states \cite{vbell1,vbell2,liprl}, three-qubit
pure states \cite{chenjingling,choudhary}, and general two-qubit quantum states \cite{yusx}, there
are no Bell inequalities yet that can be violated by all the
entangled quantum states, although it is shown recently that any entangled
multipartite pure states should violate a Bell inequality
\cite{yu}. Thus it is of great importance to find more effective Bell type
inequalities to detect the quantum entanglement.

In this paper, we study Bell inequalities for both pure and mixed multi-qubit systems.
We propose a series of Bell inequalities for any $N$-qubit states ($N\geq 3$), and
derive the formulas of the maximal violations
of these Bell inequalities. This gives a sufficient and necessary
condition which is also practical for any multipartite qubits
quantum states. It is shown that the Bell inequalities constructed
in this paper are independent of the WWZB inequality and Chen's Bell
inequalities constructed in (\ref{chenn}), i.e. they can detect some
entangled states which fulfill both the WWZB inequality and Chen's
Bell inequalities.

Consider an $N$-qubit quantum system and allow each part to choose
independently between two dichotomic observables $A_i$, $A_i^{'}$
for the $i$th observer, $i=1,2,...,N$. Each measurement has two
possible outcomes $1$ and $-1$. Quantum mechanically these
observables can be expressed as $A_i=\vec{a}_i\vec{\sigma}$,
$A_i^{'}=\vec{a}_i^{'}\vec{\sigma}$, where $\vec{\sigma}=(\sigma_1,
\sigma_2, \sigma_3)$ are the Pauli matrices and $\vec{a}_i$,
$\vec{a}_i^{'}$ are unit vectors, $i=1,2,\cdots,N$.

The CHSH Bell inequality for two-qubit systems is given by
\begin{eqnarray}\label{1}
|\la B_2\ra|\leq 1,
\end{eqnarray}
where the Bell operator
$B_2=\frac{1}{2}[A_1A_2+A_1^{'}A_2+A_1A_2^{'}-A_1^{'}A_2^{'}].$
In \cite{pla200340} Horodeckis have derived an elegant formula
which serves as a necessary and sufficient condition for violating
the CHSH inequality by an arbitrary mixed two qubits state.

The WWZB Bell operator is defined by
\be\label{wwzb}
B_N^{WWZB}=\frac{1}{2^N}\sum_{s_1,s_2,\cdots,s_N=\pm
1}S(s_1,s_2,\cdots,s_N)\sum_{k_1,k_2,\cdots,k_N=\pm
1}s_1^{k_1}s_2^{k_2}\cdots s_N^{k_N}\otimes_{j=1}^{N}O_j(k_j),
\ee
where $S(s_1,s_2,\cdots,s_N)$ is an arbitrary function of
$s_i(=\pm 1)$, $i=1,...,N$, taking values $\pm
1$, $O_j(1)=A_j$ and $O_j(2)=A_j^{'}$ with $k_j = 1, 2.$ It is shown
in \cite{wwzb1,wwzb2} that local realism requires $|\la B_N \ra|\leq
1.$ The MABK inequality is recovered by taking
$S(s_1,s_2,\cdots,s_N)=\sqrt{2}\cos[(s_1+s_2+\cdots+s_N-N+1)/\frac{\pi}{4}]$
in (\ref{wwzb}). In \cite{wwzb1,wwzb2} the authors also derived a
necessary and sufficient condition of violation of this inequality
for an arbitrary N-qubit mixed state, generalizing two-qubit results
in \cite{pla200340}. However, when using the results to obtain the
maximal violation of the WWZB inequality, one has to select a proper
set of local coordinate systems and a proper set of unit vectors,
which makes the approach less operational.

Employing an inductive method from the $(N-1)$-partite WWZB Bell
inequality to the $N$-partite inequality, a family of Bell
inequalities was presented in \cite{chenka}. The Bell operator is
defined by \be\label{chenn}
B_N=B_{N-1}^{WWZB}\otimes\frac{1}{2}(A_N+A_N^{'})+I_{N-1}\otimes\frac{1}{2}(A_N-A_N^{'}),
\ee where $B_{N-1}^{WWZB}$ represents the normal WWZB Bell operators
defined in (\ref{wwzb}), $I_{N-1}$ be the identity operators acting
on first $(N-1)$ qubits. Such Bell operators yield the violation of
the Bell inequality for the generalized GHZ states,
$|\psi\ra=\cos\alpha|00\cdots 0\ra+\sin \alpha|11\cdots 1\ra$, in
the whole parameter region of $\alpha$ such that $\cos\alpha\neq 0$
and $\sin\alpha\neq 0$, and for any number of qubits, thus
overcoming the drawback of the WWZB inequality. In the three-qubit
case, one can construct three different Bell operators from $B_{2}$
by using the approach of (\ref{chenn}). The corresponding three Bell
inequalities can distinguish full separability, detailed partial
separability and true entanglement \cite{sunbz}. However, the
maximal violation of this Bell inequality is unknown for a generally
given three-qubit state.

We start with constructing a set of new Bell inequalities for any
$N$-qubit quantum systems by iteration. First consider the case
$N=3$. As a two-qubit CHSH Bell operator $\mathcal{B}_{2}$ can act
on two of the three qubits in three different ways, we can have
three Bell operators, \be
\mathcal{B}_3^{i}=(\mathcal{B}_{2})^i\otimes\frac{1}{2}(A_i+A_i^{'})+(I_2)^i\otimes\frac{1}{2}(A_i-A_i^{'}),
~~~~i=1,2,3, \ee where $(\mathcal{B}_{2})^i$ and $(I_2)^i$ are the
two-qubit CHSH Bell operator and the identity operator acting on the
two qubits except for the $i$th one. For $N\geq 4$, the Bell
operators can be similarly obtained, \be\label{newbel}
\mathcal{B}_N^{(i-1)\frac{(N-1)!}{2}+j}=(\mathcal{B}_{N-1}^j)^i
\otimes\frac{1}{2}(A_i+A_i^{'})+(I_{N-1})^i\otimes\frac{1}{2}(A_i-A_i^{'}),
\ee with $i=1, 2, \cdots, N$ and $j=1, 2, \cdots, \frac{(N-1)!}{2}$.
Here $(\mathcal{B}_{N-1}^j)^i$ denotes the $j$th Bell operator
acting on the $(N-1)$ qubits except for the $i$th one. $(I_{N-1})^i$
stands for the identity operator acting on the $(N-1)$ qubits except
for the $i$th one. There are totally $\frac{N!}{2}$ Bell operators.

{\bf{Theorem 1:}} If a local realistic description is assumed, the
following inequalities must hold,
\be\label{thm1}
|\la\mathcal{B}_N^{k}\ra| \leq 1, \ee where $k\in\{1, 2, \cdots,
\frac{N!}{2}\}$.

{\bf{Proof:}} We prove the theorem by induction.
Note that for two qubits systems, local realism
requires that $|\la B_2\ra|\leq 1$ as shown in $(\ref{1})$.
Assume that a local realistic model has lead to
$|\la\mathcal{B}_{N-1}^{k}\ra| \leq 1$ with $k\in\{1, 2, \cdots,
\frac{(N-1)!}{2}\}$. We consider the N-partite systems.
If $A_i$ and $A_i^{'}$ are specified by
some local parameters each having two possible outcomes $-1$ and
$1$, one has either $|A_i+A_i^{'}|=2$ and $|A_i-A_i^{'}|=0$, or vice
versa. For any $k\in\{1, 2, \cdots, \frac{N!}{2}\}$,
from (\ref{newbel}) we have that
\begin{eqnarray*}
|\la\mathcal{B}_N^{k}\ra|&=&|\la\mathcal{B}_N^{(i-1)\frac{(N-1)!}{2}+j}\ra|
=|\la(\mathcal{B}_{N-1}^j)^i
\otimes\frac{1}{2}(A_i+A_i^{'})+(I_{N-1})^i\otimes\frac{1}{2}(A_i-A_i^{'})\ra|\\
&\leq& |\la(\mathcal{B}_{N-1}^j)^i\ra|
\otimes\frac{1}{2}|\la(A_i+A_i^{'})\ra|+\frac{1}{2}|\la(A_i-A_i^{'})\ra|
\leq 1.
\end{eqnarray*}
$\hfill\Box$

It is shown in \cite{terhal} that violation of a Bell inequality
gives rise formally to a kind of entanglement witness. Moreover, the separability
criterion and the existence of a description of the state by a local hidden variable
theory will become equivalent when one restricts the set of local
hidden variable theories to the domain of quantum mechanics. Thus
one can use the Bell inequalities as the separability criteria to
detect quantum entanglement. We remark that any $N$-qubit fully
separable states also satisfy the inequality (\ref{thm1}). For
$N\geq 4$, the operator $\mathcal{B}_{N-1}^i$ is derived from
$\mathcal{B}_{N-2}^i$. Thus $\mathcal{B}_N^{i}$ are different from
the Bell operators in \cite{chenka} where $\mathcal{B}_{N-1}^i$ is
the Bell operator in the WWZB inequality. The following example will
show that our Bell inequalities in (\ref{thm1}) are independent from
the WWZB inequalities and that in \cite{chenka}, and our new Bell
inequalities can detect entanglement better than they can.

{\bf{Example}} Consider a four-qubit pure state
$|\psi\ra=|\phi\ra\otimes|0\ra$, where
$|\phi\ra=\cos{\alpha}|000\ra+\sin{\alpha}|111\ra$, $\alpha\in
[0,\frac{\pi}{12}]$. It has been proved \cite{210402} that for
$\sin{2\alpha}\leq\frac{1}{2}$ (i.e. $\alpha\in
[0,\frac{\pi}{12}]$), the WWZB Bell inequalities cannot be violated
by the generalized GHZ state $|\phi\ra$. According to the result in
\cite{210402}, the WWZB inequalities operator $B_4^{WWZB}$ and the
Bell operator $B_4$ in \cite{chenka} satisfy the following
relations, \be |\la\psi|B_4^{WWZB}|\psi\ra|\leq|\la\phi|
B_3^{WWZB}|\phi\ra|\leq 1, \ee \be |\la\psi|
B_4|\psi\ra|\leq|\la\phi| B_3^{WWZB}|\phi\ra|\leq 1. \ee Therefore
both the WWZB Bell inequalities and the inequalities in
\cite{chenka} can not detect entanglement of $|\psi\ra$.

Nevertheless the mean values of the Bell operator
$\mathcal{B}_4^{12}$ in (\ref{thm1}) is
$\sqrt{2\sin^2{2\alpha}+\cos^2{2\alpha}}$ which is always larger
than 1 as long as $|\phi\ra$ is not separable. Therefore the
entanglement is detected by our Bell inequality (\ref{thm1}).

To identify the non-local properties of a quantum state, it is important to find the necessary and sufficient conditions for the quantum state to violate the Bell inequality. Now we
investigate the maximal violation of the Bell inequalities
(\ref{thm1}). We first consider the $N=3$ case. In this situation,
(\ref{newbel}) gives three operators,
\begin{eqnarray}\label{31}\mathcal{B}_3^{1}&=&(\mathcal{B}_2)^{1}\otimes \frac{1}{2}(A_1+A_1^{'})
+(I_2)^1\otimes\frac{1}{2}(A_1-A_1^{'}),\end{eqnarray}
\begin{eqnarray}\label{32}\mathcal{B}_3^{2}&=&(\mathcal{B}_2)^{2}\otimes \frac{1}{2}(A_2+A_2^{'})
+(I_2)^2\otimes\frac{1}{2}(A_2-A_2^{'}),\end{eqnarray}
\begin{eqnarray}\label{33}\mathcal{B}_3^{3}&=&(\mathcal{B}_2)^{3}\otimes \frac{1}{2}(A_3+A_3^{'})
+(I_2)^3\otimes\frac{1}{2}(A_3-A_3^{'}),\end{eqnarray} where
$(\mathcal{B}_2)^{1}=\frac{1}{2}(A_2A_3+A_2^{'} A_3+A_2
A_3^{'}-A_2^{'} A_3^{'})$,
$(\mathcal{B}_2)^{2}=\frac{1}{2}(A_1A_3+A_1^{'} A_3+A_1
A_3^{'}-A_1^{'} A_3^{'})$ and
$(\mathcal{B}_2)^{3}=\frac{1}{2}(A_1A_2+A_1^{'} A_2+A_1
A_2^{'}-A_1^{'} A_2^{'})$. Let $\rho$ be a general three-qubit
state, \be\label{rho}
\rho=\sum_{i,j,k=0}^3T_{ijk}\sigma_i\sigma_j\sigma_k, \ee where
$\sigma_0=I_2$ is the $2\times 2$ identity matrix, $\sigma_i$ are
the Pauli matrices, and \be\label{t}T_{ijk}=\frac{1}{8} Tr(\rho
\sigma_i\sigma_j\sigma_k).\ee

{\bf{Theorem 2:}} The maximum of the mean values of the Bell
operators in (\ref{31}), (\ref{32}) and (\ref{33}) satisfy the following relations,
\begin{eqnarray}\label{nnv1}
\max|\la\mathcal{B}_3^{1} \ra|
=8\max\{\lambda_1^1(\vec{b}_3)+\lambda_2^1(\vec{b}_3)+
||\vec{T}^1_{00}||^2-\la \vec{b}_3, \vec{T}^1_{00}
\ra^2\}^{\frac{1}{2}},\\[2mm]\label{nnv2}
\max|\la\mathcal{B}_3^{2} \ra|
=8\max\{\lambda_1^2(\vec{b}_3)+\lambda_2^2(\vec{b}_3)+
||\vec{T}^2_{00}||^2-\la \vec{b}_3, \vec{T}^2_{00}
\ra^2\}^{\frac{1}{2}},\\[2mm]\label{nnv3}
\max|\la\mathcal{B}_3^{3} \ra|
=8\max\{\lambda_1^3(\vec{b}_3)+\lambda_2^3(\vec{b}_3)+
||\vec{T}^3_{00}||^2-\la \vec{b}_3, \vec{T}^3_{00}
\ra^2\}^{\frac{1}{2}},
\end{eqnarray}
where $\la.,.\ra$ denotes the inner product of two vectors,
$||\vec{x}||$ stands for the norm of vector $\vec{x}$. The maximums
on the right of (\ref{nnv1}), (\ref{nnv2}) and (\ref{nnv3}) are
taken over all the unit vectors $\vec{b}_3$. Given a three-qubit
state $\rho$, one can compute $T_{ijk}$ by using the
formula in (\ref{t}). Then $\lambda_1^i(\vec{b}_3)$ and
$\lambda_2^i(\vec{b}_3)$ are defined to be the two greater
eigenvalues of the matrix $M_i^{\dag}M_i$ with
$M_i=\sum_{k=1}^3b_3^kT_k^i$, $i=1,2,3$, with respect to the three
Bell operators in (\ref{31}), (\ref{32}) and (\ref{33}). Here
$T_k^l$, $l=1,2,3$, are matrices with entries given by
$(T_k^1)_{ij}=T_{kij}$, $(T_k^2)_{ij}=T_{ikj}$ and
$(T_k^3)_{ij}=T_{ijk}$. $\vec{T}_{00}^m$, $m=1,2,3$ are defined to be
vectors with entries $(\vec{T}_{00}^1)_k=T_{k00}$,
$(\vec{T}_{00}^2)_k=T_{0k0}$ and $(\vec{T}_{00}^3)_k=T_{00k}$.

{\bf{Proof:}}
We take (\ref{33}) as an example to show how to calculate the
maximal violation. The maximal violation for the Bell operators (\ref{31}) and
(\ref{32}) can be computed similarly. A direct computation shows
that
\begin{eqnarray}\label{cal}\mathcal{B}_3^{3}&=&\frac{1}{4}[(A_1+A_1^{'})A_2+(A_1-A_1^{'})A_2^{'}](A_3+A_3^{'})+(I_2)^3\otimes\frac{1}{2}(A_3-A_3^{'})\nonumber\\
&=&\frac{1}{4}(A_1+A_1^{'})A_2(A_3+A_3^{'})+\frac{1}{4}(A_1-A_1^{'})A_2^{'}(A_3+A_3^{'})+(I_2)^3\otimes\frac{1}{2}(A_3-A_3^{'})\nonumber\\
&=&\frac{1}{4}\sum_{i,j,k=1}^3[a_1^i+(a_1^{'})^i]a_2^j[a_3^k+(a_3^{'})^k]\sigma_i\sigma_j\sigma_k
+\frac{1}{4}\sum_{i,j,k=1}^3[a_1^i-(a_1^{'})^i](a_2^{'})^j[a_3^k+(a_3^{'})^k]\sigma_i\sigma_j\sigma_k\nonumber\\
&&+\frac{1}{2}\sum_{k=1}^3[a_3^k-(a_3^{'})^k]I_4\otimes\sigma_k.
\end{eqnarray}

For any given unit vectors $\vec{a}_1$ and $\vec{a^{'}}_1$,
there always exist a pair of unit and mutually orthogonal vectors
$\vec{b}_1$, $\vec{b^{'}}_1$ and $\theta\in [0,\frac{\pi}{2}]$ such that
\be\label{use1}
\vec{a}_1+\vec{a^{'}}_1=2\cos{\theta}\,\vec{b}_1,~~~ \vec{a}_1-\vec{a^{'}}_1=2\sin{\theta}\,\vec{b^{'}}_1.
\ee
Similarly for $\vec{a}_3$ and $\vec{a^{'}}_3$, we have
\be\label{use2}
\vec{a}_3+\vec{a^{'}}_3=2\cos{\phi}\,\vec{b}_3,~~~
\vec{a}_3-\vec{a^{'}}_3=2\sin{\phi}\,\vec{b^{'}}_3,
\ee
where $\vec{b}_3$ and $\vec{b^{'}}_3$ are orthogonal vectors with unit norm and $\phi\in [0,\frac{\pi}{2}]$.

By inserting (\ref{rho}) into (\ref{cal}) we get the mean
value of the Bell operator (\ref{33}),
\begin{eqnarray*}
\la\mathcal{B}_3^{3} \ra&=& Tr(\rho\mathcal{B}_3^{3})\\
&=&\frac{1}{4}\sum_{i,j,k=1}^3[a_1^i+(a_1^{'})^i]a_2^j[a_3^k+(a_3^{'})^k]T_{ijk}
Tr(\sigma_i^2\sigma_j^2\sigma_k^2)\nonumber\\
&&+\frac{1}{4}\sum_{i,j,k=1}^3[a_1^i-(a_1^{'})^i](a_2^{'})^j[a_3^k+(a_3^{'})^k]T_{ijk}
Tr(\sigma_i^2\sigma_j^2\sigma_k^2)\nonumber\\
&&+\frac{1}{2}\sum_{k=1}^3[a_3^k-(a_3^{'})^k]T_{00k}Tr(I_4^2\otimes\sigma_k^2)\nonumber\\
&=&8\sum_{i,j,k=1}^3b_1^ib_3^ka_2^jT_{ijk}\cos{\theta}\cos{\phi}
+8\sum_{i,j,k=1}^3(b_1^{'})^ib_3^k(a_2^{'})^jT_{ijk}\sin{\theta}\cos{\phi}
+4\sum_{k=1}^3(b_3^{'})^kT_{00k}\sin{\phi}.\end{eqnarray*}

Let $T_k^3, k=1,2,3$, be the matrux with entries given by
$(T_k^3)_{ij}=T_{ijk}$ and $\vec{T}^3_{00}$ a vector with
components $(\vec{T}^3_{00})_k=T_{00k}$. The maximal mean value
of the Bell operator (\ref{33}) can be written as
\begin{eqnarray}\label{nv}
\max\la\mathcal{B}_3^{3} \ra&=& 8\max[\la\vec{b}_1,\sum_{k=1}^3b_3^kT^3_k\vec{a}_2\ra\cos{\theta}\cos{\phi}
+\la\vec{b}_1^{'},\sum_{k=1}^3b_3^kT^3_k\vec{a}_2^{'}\ra\sin{\theta}\cos{\phi}+\la \vec{b}_3^{'}, \vec{T}^3_{00} \ra\sin{\phi}]\nonumber\\
&=&8\max\{[\la\vec{b}_1,\sum_{k=1}^3b_3^kT^3_k\vec{a}_2\ra\cos{\theta}+\la\vec{b}_1^{'},\sum_{k=1}^3b_3^kT^3_k\vec{a}_2^{'}\ra\sin{\theta}]^2
+\la \vec{b}_3^{'}, \vec{T}^3_{00} \ra^2\}^{\frac{1}{2}}\nonumber\\
&=&8\max\{[\la\vec{b}_1,\sum_{k=1}^3b_3^kT^3_k\vec{a}_2\ra\cos{\theta}+\la\vec{b}_1^{'},\sum_{k=1}^3b_3^kT^3_k\vec{a}_2^{'}\ra\sin{\theta}]^2+
||\vec{T}^3_{00}||^2-\la \vec{b}_3, \vec{T}^3_{00} \ra^2\}^{\frac{1}{2}}\nonumber\\
&=&8\max\{\lambda_1^3(\vec{b}_3)+\lambda_2^3(\vec{b}_3)+
||\vec{T}^3_{00}||^2-\la \vec{b}_3, \vec{T}^3_{00}
\ra^2\}^{\frac{1}{2}},
\end{eqnarray}
which proves (\ref{nnv3}). In (\ref{nv}) we have used the fact that
the maximum of $x\cos{\theta}+y\sin{\theta}$ taking over all $\theta$
is $\sqrt{x^2+y^2}$. Formulae (\ref{nnv1}) and (\ref{nnv2})
can be similarly proven. $\hfill\Box$

{\bf{Remark:}} According to the symmetry of the operator $\mathcal{B}_3^{3}$,
the equation (\ref{nv}) also provides the minimum of the operator
(\ref{33}), achieved by $-\mathcal{B}_3^{3}$.

\begin{figure}[tbp]
\begin{center}
\resizebox{10cm}{!}{\includegraphics{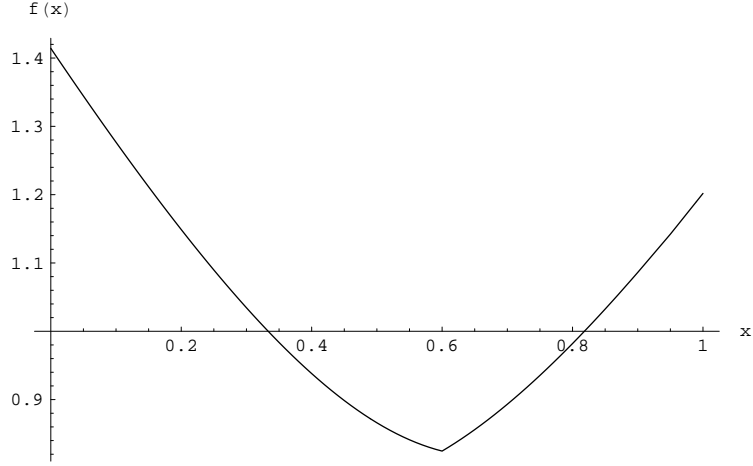}}
\end{center}
\caption{The maximal mean value of the operator (\ref{33}) for the
mixed state $\rho$ in (\ref{state}). $f(x)$ stands for the
maximal mean value and $x$ is the parameter in $\rho$.\label{fig1}}
\end{figure}

Since $\vec{b}_3$ is a three dimensional real unit vector, one can
always calculate the exact value of the maximum for any given three
qubits quantum state. For example, for the generalized three-qubit
GHZ state, $|GHZ\ra=\cos{\alpha}|000\ra+\sin{\alpha}|111\ra$, by
selecting some proper direction of the measurement operators, i.e.
$\vec{a}_i$s and $(\vec{a}^{'})_i$s, the maximal mean value of the
Bell operator in (\ref{33}) is shown to be \cite{chenka},
$\sqrt{2\sin^2{2\alpha}+\cos^2{2\alpha}}$. From our formulae in
Theorem 2 one can show that the result are  in accord with that in
\cite{chenka}. For three-qubit $W$ state,
$|W\ra=\frac{1}{\sqrt{3}}(|100\ra+|010\ra+|001\ra)$, our mean value
is 1.202, which is also in agreement with that in \cite{chenka}.
However, our method can be also used to calculate the mean value of
the Bell operators in (\ref{31}), (\ref{32}) and (\ref{33}) for any
three qubits quantum states. For instance, we consider the mixture
of $|W\ra$ and $|GHZ\ra$, \be\label{state}\rho=x|W\ra\la
W|+(1-x)|GHZ\ra\la GHZ|,\ee where $0\leq x\leq 1$. We have the
maximal mean value of the Bell operator in (\ref{33}),see
Fig.\ref{fig1}, where $f(x)$ stands for the maximal mean value of
the operator (\ref{33}) for the mixed state $\rho$. For $0\leq x\leq
0.33$ and $0.82\leq x\leq1$, $f(x)>1$ and $\rho$ is detected to be
entangled.

For four-qubit systems, the Bell operators (\ref{newbel}) have
four different forms. We take $\mathcal{B}_4^{12}$ as an example to
investigate the maximal violation of the corresponding Bell
inequality. Note that
\begin{eqnarray}\label{cal}\mathcal{B}_4^{12}&=&(\mathcal{B}_3^{3})^4\otimes \frac{1}{2}(A_4+A_4^{'})+(I_3)^4\otimes\frac{1}{2}(A_4-A_4^{'})\nonumber\\
&=&\frac{1}{8}(A_1+A_1^{'})A_2(A_3+A_3^{'})(A_4+A_4^{'})+\frac{1}{8}(A_1-A_1^{'})A_2^{'}(A_3+A_3^{'})(A_4+A_4^{'})\nonumber\\
&&+\frac{1}{4}I_4\otimes(A_3-A_3^{'})(A_4+A_4^{'})+\frac{1}{2}I_6\otimes(A_4-A_4^{'})\nonumber\\
&=&\frac{1}{8}\sum_{i,j,k,l=1}^3[a_1^i+(a_1^{'})^i]a_2^j[a_3^k+(a_3^{'})^k][a_4^l+(a_4^{'})^l]\sigma_i\sigma_j\sigma_k\sigma_l\nonumber\\
&&+\frac{1}{8}\sum_{i,j,k,l=1}^3[a_1^i-(a_1^{'})^i](a_2^{'})^j[a_3^k+(a_3^{'})^k][a_4^l+(a_4^{'})^l]\sigma_i\sigma_j\sigma_k\sigma_l\nonumber\\
&&+\frac{1}{4}\sum_{k,l=1}^3[a_3^k-(a_3^{'})^k][a_4^l+(a_4^{'})^l]I_4\otimes\sigma_k\sigma_l
+\frac{1}{2}\sum_{l=1}^3[a_4^l-(a_4^{'})^l]I_6\otimes\sigma_l.
\end{eqnarray}
Let $\rho$ be a general four-qubit quantum state,
\be\label{rho4}
\rho=\sum_{i,j,k,l=0}^3T_{ijkl}\sigma_i\sigma_j\sigma_k\sigma_l,
\ee
with $T_{ijkl}=\frac{1}{2^4} Tr(\rho
\sigma_i\sigma_j\sigma_k\sigma_l)$. The mean value of
$\mathcal{B}_4^{12}$ can be derived by the following deduction.
\begin{eqnarray*} \la\mathcal{B}_4^{12} \ra
&=&\frac{1}{8}\sum_{i,j,k,l=1}^3[a_1^i+(a_1^{'})^i]a_2^j[a_3^k+(a_3^{'})^k][a_4^l+(a_4^{'})^l]T_{ijkl}Tr(\sigma_i^2\sigma_j^2\sigma_k^2\sigma_l^2)\nonumber\\
&&+\frac{1}{8}\sum_{i,j,k,l=1}^3[a_1^i-(a_1^{'})^i](a_2^{'})^j[a_3^k+(a_3^{'})^k][a_4^l+(a_4^{'})^l]T_{ijkl}Tr(\sigma_i^2\sigma_j^2\sigma_k^2\sigma_l^2)\\
&&+\frac{1}{4}\sum_{k,l=1}^3[a_3^k-(a_3^{'})^k][a_4^l+(a_4^{'})^l]T_{00kl}Tr(I_4\otimes\sigma_k^2\sigma_l^2)\nonumber\\
&&+\frac{1}{2}\sum_{l=1}^3[a_4^l-(a_4^{'})^l]T_{000l}Tr(I_6\otimes\sigma_l^2)\nonumber\\
&=&2^4\sum_{i,j,k,l=1}^3b_1^ia_2^jb_3^kb_4^lT_{ijkl}\cos{\alpha_1}\cos{\alpha_3}\cos{\alpha_4}\nonumber\\
&&+2^4\sum_{i,j,k,l=1}^3(b_1^{'})^i(a_2^{'})^jb_3^kb_4^lT_{ijkl}\sin{\alpha_1}\cos{\alpha_3}\cos{\alpha_4}\nonumber\\
&&+2^4\sum_{k,l=1}^3(b_3^{'})^kb_4^lT_{00kl}\sin{\alpha_3}\cos{\alpha_4}+2^4\sum_{l=1}^3(b_4^{'})^lT_{000l}\sin{\alpha_4},
\end{eqnarray*}
where we have assumed that
$\vec{a}_i+\vec{a^{'}}_i=2\cos{\alpha_i}\,\vec{b}_i$,
$\vec{a}_i-\vec{a^{'}}_i=2\sin{\alpha_i}\,\vec{b^{'}}_i$,
$\alpha_i\in[0,\frac{\pi}{2}]$.

The maximum of the mean value can be derived to be
\begin{eqnarray*}\label{nv4} \max\la\mathcal{B}_4^{12} \ra
&=&2^4\max[\la\vec{b}_1,\sum_{k,l=1}^3b_3^kb_4^lT^{12}_{kl}\vec{a}_2\ra\cos{\alpha_1}\cos{\alpha_3}\cos{\alpha_4}\\
&&+\la\vec{b}_1^{'},\sum_{k,l=1}^3b_3^kb_4^lT^{12}_{kl}\vec{a}_2^{'}\ra\sin{\alpha_1}\cos{\alpha_3}\cos{\alpha_4}\\
&&+\la \vec{b}_3^{'}, T^{12}_{00}\vec{b}_4 \ra\sin{\alpha_3}\cos{\alpha_4}]+\la \vec{b}_4^{'}, \vec{T}^{12}_{000} \ra\sin{\alpha_4}]\\
&=&2^4\max\{\la\vec{b}_1,\sum_{k,l=1}^3b_3^kb_4^lT^{12}_{kl}\vec{a}_2\ra^2
+\la\vec{b}_1^{'},\sum_{k,l=1}^3b_3^kb_4^lT^{12}_{kl}\vec{a}_2^{'}\ra^2
+\la \vec{b}_3^{'}, T^{12}_{00}\vec{b}_4 \ra^2+\la \vec{b}_4^{'}, \vec{T}^{12}_{000}\ra^2\}^{\frac{1}{2}}\\
&=&2^4\max\{\lambda_1^{12}(\vec{b}_3\vec{b}_4)+\lambda_2^{12}(\vec{b}_3\vec{b}_4)+
||T^{12}_{00}\vec{b}_4||^2-\la
\vec{b}_3,T_{00}^{12}\vec{b}_4\ra^2+||\vec{T}^{12}_{000}||^2-\la
\vec{b}_4, \vec{T}^{12}_{000} \ra^2\}^{\frac{1}{2}},\end{eqnarray*}
where $\lambda_1^{12}(\vec{b}_3\vec{b}_4)$ and
$\lambda_2^{12}(\vec{b}_3\vec{b}_4)$ are the two greater eigenvalues
of the matrix $(M^{12})^{\dag}M^{12}$,
$M^{12}=\sum_{k,l=1}^3b_3^kb_4^lT_{kl}^{12}; T_{kl}^{12}$ stand for
the matrices with entries $(T_{kl}^{12})_{ij}=T_{ijkl}$ with
$i,j,k,l=1,2,3$; $T_{00}^{12}$ is a matrix with entries
$(T_{00}^{12})_{kl}=T_{00kl}$, and $\vec{T}^{12}_{000}$ is a vector
with components $(T^{12}_{000})_l=T_{000l}$, $l=1,2,3$. The maximum
in the last equation is taken over all the unit vectors $\vec{b}_3$
and $\vec{b}_4$.

In terms of the analysis above, for four-qubit systems we have the
following Theorem.

{\bf{Theorem 3:}} The maximum of the mean values of the Bell
operators in (\ref{newbel}) for four qubits systems are given by the following formula:
\begin{eqnarray}\label{nv4all} \max|\la\mathcal{B}_4^{m} \ra|
=2^4\max\{\lambda_1^m(\vec{b}_3\vec{b}_4)+\lambda_2^m(\vec{b}_3\vec{b}_4)+
||T^m_{00}\vec{b}_4||^2-\la
\vec{b}_3,T_{00}^m\vec{b}_4\ra^2+||\vec{T}^m_{000}||^2-\la
\vec{b}_4, \vec{T}^m_{000} \ra^2\}^{\frac{1}{2}}.
\end{eqnarray}
The maximums on the right side are taken over all the unit vectors
$\vec{b}_3$ and $\vec{b}_4$. Here $\lambda_1^m(\vec{b}_3\vec{b}_4)$
and $\lambda_2^m(\vec{b}_3\vec{b}_4)$ are the two greater
eigenvalues of the matrix $(M^m)^{\dag}M^m$,
$M^m=\sum_{k,l=1}^3b_3^kb_4^lT_{kl}^m$, $m=1, 2, \cdots,
\frac{N!}{2}$; $T_{kl}^m$ are the matrices with entries
$(T_{kl}^1)_{ij}=T_{lkij}$, $(T_{kl}^2)_{ij}=T_{likj}$,
$(T_{kl}^3)_{ij}=T_{lijk}$, $(T_{kl}^4)_{ij}=T_{klij}$,
$(T_{kl}^5)_{ij}=T_{ilkj}$, $(T_{kl}^6)_{ij}=T_{iljk}$,
$(T_{kl}^7)_{ij}=T_{kilj}$, $(T_{kl}^8)_{ij}=T_{iklj}$,
$(T_{kl}^9)_{ij}=T_{ijlk}$, $(T_{kl}^{10})_{ij}=T_{kijl}$,
$(T_{kl}^{11})_{ij}=T_{ikjl}$ and $(T_{kl}^{12})_{ij}=T_{ijkl}$, $i,
j=1, 2, 3$ and $k, l=0, 1, 2, 3$; $\vec{T}^m_{000}$ stand for
the vectors with components $(T^i_{000})_x=T_{x000}$, $(T^j_{000})_x=T_{0x00}$, $(T^k_{000})_x=T_{00x0}$,
$(T^l_{000})_x=T_{000x}$, $i=1,2,3$, $j=4,5,6$, $k=7,8,9$,
$l=10,11,12$ and $x=1,2,3$.

As an example, consider the 4-qubit W state
$|W\ra=\frac{1}{2}(|1000\ra+|0100\ra+|0010\ra+|0001\ra)$, by using
the formula (\ref{nv4all}) one gets the maximal mean value
$\max|\la\mathcal{B}_4^{12}\ra|=1.118$. For the mixed state
$\rho=\frac{x}{16}I+(1-x)|W\ra\la W|$, entanglement can be detected
by (\ref{nv4all}) for $0 \leq x\leq 0.106$.

Generally, for any N-qubit quantum state, the maximal mean values of
the Bell operators in (\ref{newbel}) can be calculated similarly by
using our approach above. For example, the maximal mean value of
$\mathcal{B}_N^{\frac{N!}{2}}$ can be expressed as
\begin{eqnarray}\label{nvn} \max|\la\mathcal{B}_N^{\frac{N!}{2}} \ra|
&=&2^N\max\{\lambda_1^m(\vec{b}_3\cdots\vec{b}_N)+\lambda_2^m(\vec{b}_3\cdots\vec{b}_N)+
||\vec{T}_{45\cdots N}||^2-\la \vec{b}_3,\vec{T}_{45\cdots
N}\ra^2\nonumber\\
&+&||\vec{T}_{5\cdots N}||^2-\la \vec{b}_4,\vec{T}_{5\cdots
N}\ra^2+\cdots+||\vec{T}_{N}||^2-\la
\vec{b}_N,\vec{T}_{N}\ra^2\}^{\frac{1}{2}},
\end{eqnarray}
where $\lambda_1(\vec{b}_3\cdots\vec{b}_N)$ and
$\lambda_2(\vec{b}_3\cdots\vec{b}_N)$ are the two greater
eigenvalues of the matrix $M^{\dag}M$, with
$(M)_{ij}=\sum_{i_3,\cdots, i_N=1}^3b_3^{i_3}\cdots
b_N^{i_N}T_{iji_3,\cdots, i_N}$ the entries of matrix $M$;
$\vec{T}_{45\cdots N}$, $\vec{T}_{5\cdots N}$ and $\vec{T}_{N}$ are
vectors with components $(\vec{T}_{45\cdots N})_k=\sum_{i_4,\cdots,
i_N=1}^3b_4^{i_4}\cdots b_N^{i_N}T_{00ki_4,\cdots, i_N},
(\vec{T}_{5\cdots N})_k=\sum_{i_5,\cdots, i_N=1}^3b_5^{i_5}\cdots
b_N^{i_N}T_{000ki_5,\cdots, i_N}$ and $(\vec{T}_{N})_k=T_{000\cdots0
k}$ respectively. The maximum on the right side is taken over all
the unit vectors $\vec{b}_3, \vec{b}_4, \cdots, \vec{b}_N$. The
other mean values of the Bell operators in (\ref{newbel}) for
$N$-qubit states can be obtained similarly.
By expressing the unit vectors $\vec{b}_k$ as
$(\cos{\theta_k}\cos{\phi_k},\cos{\theta_k}\sin{\phi_k},\sin{\theta_k})$,
$k=3,\cdots, N$, our formulas can be used to compute the
maximal violation by searching for the maximum over all
$\theta_k$ and $\phi_k$, either analytically or numerically.

In conclusion, we have presented a series of Bell inequalities for
multipartite qubits systems. These Bell inequalities are more effective in detecting the non-local properties of quantum states that can not be described by local realistic models.
Formulas are derived to calculate the maximal
violation of the Bell inequalities for any given multiqubit states,
which gives rise to the sufficient and necessary condition for the
violation of these Bell inequalities. For a fixed multiqubit state,
one can optimize the mean values of the Bell operators over all
measurement directions. Our Bell operators only involve two
measurement settings per site, which meets the simplicity
requirements of current linear optical experiments for nonlocality
tests. Moreover, our formulas for the maximal violation of the Bell
inequalities fit for both pure and mixed states, and can be used to
improve the detection of multiqubit entanglement.

\bigskip
\noindent{\bf Acknowledgments}\, This work is supported by the NSFC
11105226 and PHR201007107.

\end{document}